\documentclass[a4paper]{jpconf}
\usepackage{graphicx}
\usepackage{cite}
\begin{document}
\title{Dynamical properties of Potts model with invisible states}

\author{Shu Tanaka$^1$ and Ryo Tamura$^2$}

\address{
$^1$Research Center for Quantum Computing, Interdisciplinary Graduate School of Science and Engineering, Kinki University.
3-4-1, Kowakae, Higashi-Osaka, Osaka 577-8502, Japan
}
\address{
$^2$Institute for Solid State Physics, University of Tokyo.
5-1-5, Kashiwanoha Kashiwa, Chiba 277-8581, Japan
}

\ead{shu-t@alice.math.kindai.ac.jp}

\begin{abstract}
We study dynamic behavior of Potts model with invisible states near the first-order phase transition temperature.
We focus on melting process starting from the perfect ordered state.
This model is regarded as a standard model to analyse nature of phase transition.
We can control the energy barrier between the ordered state and paramagnetic state without changing the symmetry which breaks at the transition point.
We calculate time-dependency of the order parameter, density of invisible state, and internal energy.
They show two-step relaxation behavior.
We also consider the relation between the characteristic melting time and characteristic scale of the energy barrier by changing the number of invisible states.
We find that characteristic melting time increases as the energy barrier enlarges in this model.
Thus, this model is regarded as a fundamental model to analyze dynamic behavior near the first-order phase transition point.
\end{abstract}

\section{Introduction}

Frustration causes many interesting static and dynamic behavior which are not observed in unfrustrated systems because of peculiar density of states\cite{Toulouse-1977,Liebmann-1986,Kawamura-1998,Diep-2005,Moessner-2000,Matsuda-2009,Tanaka-2009,Tanaka-2010a,Tanaka-2010b,Tanaka-2010c}. 
In two-dimensional frustrated systems, there have been found many nontrivial phase transitions such as order by disorder\cite{Villain-1980,Henley-1989}, reentrant phase transition\cite{Nakano-1968,Syozi-1968,Fradkin-1976,Tanaka-2005,Miyashita-2007}, topological phase transition\cite{Kawamura-1984}, and novel type of first-order phase transition. 
Recently, strange first-order phase transitions have been found in two-dimensional frustrated continuous spin systems\cite{Tamura-2008,Stoudenmire-2009,Okumura-2010,Tamura-inp}. 

In \cite{Tamura-2008}, the authors studied equilibrium properties of the classical Heisenberg model on triangular lattice with nearest neighbor ferromagnetic interaction $J_1$ and third-nearest neighbor antiferromagnetic interaction $J_3$.
They found that a first-order phase transition with threefold symmetry breaking occurs at finite temperature.
This looks a strange phase transition, since phase transition with threefold symmetry breaking is often second-order phase transition on two-dimensional lattice {\it e.g.} the three-state ferromagnetic Potts model or the three-state ferromagnetic clock model\cite{Wu-1982}. 
After this study, similar nature of first-order phase transition have been found by a number of researchers\cite{Stoudenmire-2009,Okumura-2010,Tamura-inp}. 
Stoudenmire {\it et al.} found a first-order phase transition with breaking of threefold symmetry in $J_1-J_3$ model with biquadratic interaction on triangular lattice.
Okumura {\it et al.} also found similar first-order phase transition in frustrated $J_1-J_2$ model on hexagonal lattice. 
Such a first-order phase transition sometimes takes place in two-dimensional frustrated systems with a number of competed interactions.
It is an open problem why such a first-order phase transision appears in some two-dimensional frustrated systems.

To consider this problem, we constructed a model which exhibits a first-order phase transition with threefold symmetry breaking by introducing new kind of parameter into the standard ferromagnetic three-state Potts model\cite{Tamura-2010,Tanaka-2010d}.
We introduced the invisible states which does not contribute to the internal energy.
The ground state of the Potts model with invisible state and its degeneracy are the same as that of the standard Potts model.
We found that the first-order phase transition with threefold symmetry breaking occurs by just adding the invisible states into the standard three-state ferromagnetic Potts model.
The standard Potts model is regarded as a fundamental model to consider the properties of phase transition in statistical physics\cite{Potts-1952,Wu-1982}.
It is believed that the Potts model with invisible states is a potential model which can clarify inherent nature of first-order phase transition with breaking of threefold symmetry on two-dimensional lattice.

Our aim of the present study is to clarify dynamic properties of Potts model with invisible states.
It is interesting topic in statistical physics to study dynamical nature of the systems which exhibit a first-order phase transition, since there are some similar points between these systems and glassy systems\cite{Krzakala-2010a,Krzakala-2010b}.
It is expected that some dynamic nature in glassy systems can be explained in terms of dynamic behavior of the systems which have a first-order phase transition.
Potts model with invisible states can be considered a potential model to study dynamical nature of first-order phase transition.
This is because we can control the energy barrier between the ordered state and paramagnetic state without changing the symmetry which breaks at the transition temperature by modulating the number of invisible states.
Especially, we focus on melting process in this model.
In this paper, we study dynamics of the Potts model with invisible states at fixed temperatures which are above the transition temperature.

\section{Model}

We consider the ferromagnetic Potts model with invisible states on square lattice.
The Hamiltonian of this model is 
\begin{eqnarray}
 &&{\cal H} = -J \sum_{\langle i,j \rangle} \delta_{\sigma_i,\sigma_j} \sum_{\alpha=1}^q \delta_{\sigma_i,\alpha},\,\,\,\, (J>0)\\
 &&\sigma_i = 1, \cdots, q, q+1, \cdots, q+r,
\end{eqnarray}
where $\langle i,j \rangle$ denotes the nearest neighbor pairs on square lattice.
In this paper we take $J$ as an energy unit.
If and only if $1 \le \sigma_i = \sigma_j \le q$, the interaction works.
Obviously this model for $r=0$ corresponds to the standard $q$-state ferromagnetic Potts model.
We call the state in $1 \le \sigma_i \le q$ ``colored state'' and the state in $q+1 \le \sigma_i \le q+r$ ``invisible state'' from now on.
Hereafter we call this model ($q$,$r$)-state Potts model.
The invisible state does not affect the internal energy.
It should be noted that the number of ground states of the ($q$,$r$)-state Potts model is the same as that of the standard ferromagnetic $q$-state Potts model.
Then the ($q$,$r$)-state Potts model exhibits a phase transition with $q$-fold symmetry breaking.
To change the number of invisible states $r$ corresponds to changing form of the density of states.
The invisible states contributes the entropy as $\log r$.
Then, it is expected that the order of phase transition can be changed by adding the invisible states.
Actually, a first-order phase transition with $q$-fold symmetry breaking occurs for large enough the number of invisible states $r$ even for $q=2,3,$ and $4$ on two-dimensional lattice\cite{Tamura-2010}.
As the number of the invisible states $r$ increases, the latent heat increases and the transition temperature decreases\cite{Tanaka-2010d}.

\section{Result}

In this paper, we study dynamics of the ($q$,$r$)-state Potts model on square lattice whose size is $N=128\times 128$ at fixed temperatures which are above the transition temperature following \cite{Krzakala-2010b}.
We impose periodic boundary condition.
We prepare independent $1024$ samples for obtaining data with high accuracy.
The initial state is set to be perfect ordered state such as $\sigma_i = 1$ for all $i$.
We adopt single-spin-flip Metropolis type of Monte Carlo method as the time-evolution rule.

To consider dynamic behavior of the ($q$,$r$)-state Potts model, we define the time-dependent order parameter $m(t)$, density of invisible states $\rho_{\rm inv}(t)$, and internal energy $e(t)$ as follows:
\begin{eqnarray}
 &&m(t) = \frac{(q+r)\sum_{i=1}^N \left[ \delta_{\sigma_i(t),1} - \frac{1}{q+r}\right]}{N(q+r-1)},\\
 &&\rho_{\rm inv}(t) = \frac{1}{N} \sum_{i=1}^N \sum_{\alpha=q+1}^{q+r} \delta_{\sigma_i(t),\alpha},\\
 &&e(t) = -\frac{J}{N} \sum_{\langle i,j \rangle} \delta_{\sigma_i(t),\sigma_j(t)} \sum_{\alpha=1}^q \delta_{\sigma_i(t),\alpha},
\end{eqnarray}
where $t$ denotes Monte Carlo step.
\begin{figure}[b]
 \begin{center}
  \includegraphics[scale=0.65]{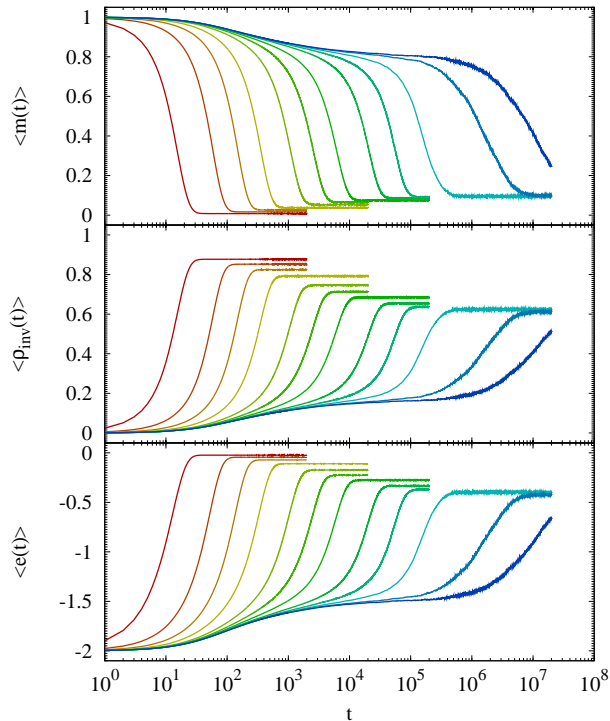}
  \caption{
  Time-evolution of the order parameter $\langle m(t)\rangle$, density of the invisible states $\langle \rho_{\rm inv}(t) \rangle$, and the internal energy $\langle e(t) \rangle$ of the ($3$,$27$)-state Potts model starting from the completely ordered state $\sigma_i(0)=1$ (for all $i$) for fixed different temperatures.
From left to right, the temperatures are $T=T_{\rm c}+0.5$, $T_{\rm c}+0.2$, $T_{\rm c}+0.1$, $T_{\rm c}+0.05$, $T_{\rm c}+0.02$, $T_{\rm c}+0.01$, $T_{\rm c}+0.005$, $T_{\rm c}+0.002$, $T_{\rm c}+0.001$, $T_{\rm c}+0.0005$, $T_{\rm c}+0.0002$, and $T_{\rm c}+0.0001$, where $T_{\rm c}=0.58513$\cite{Tanaka-2010d}.
  }
  \label{graph:timeev}
 \end{center}
\end{figure}
Figure \ref{graph:timeev} shows the time-evolution of the order parameter $\langle m(t)\rangle$, density of the invisible states $\langle \rho_{\rm inv}(t) \rangle$, and the internal energy $\langle e(t) \rangle$ of the ($3$,$27$)-state Potts model starting from the completely ordered state $\sigma_i(0)=1$ (for all $i$) for fixed different temperatures.
Here $\langle \cdot \rangle$ denotes ensemble average.
In Fig.~\ref{graph:timeev}, there are obvious two-step relaxations.
Two-step relaxation often appears in systems where a first-order phase transition takes place and also in glassy systems\cite{Tanaka-2007}.
Melting occurs after stabilization in the ordered state which corresponds to plateau region.
As the temperature approaches to the transition point, the plateau region enlarges and as a result, the characteristic melting time becomes long.

Next we consider the dynamical susceptibility to consider the characteristic melting time systematically.
The dynamical susceptibility of the time-dependent physical quantity $A(t)$ is defined as
\begin{eqnarray}
 \chi_A(t) = N \beta (\langle A(t)^2 \rangle - \langle A(t) \rangle^2),
\end{eqnarray}
where $\beta$ denotes the inverse temperature.
We calculate the dynamical susceptibility of the order parameter $\chi_m(t)$ and density of the invisible states $\chi_{\rho_{\rm inv}}(t)$.
\begin{figure}[t]
 \begin{center}
  \includegraphics[scale=0.65]{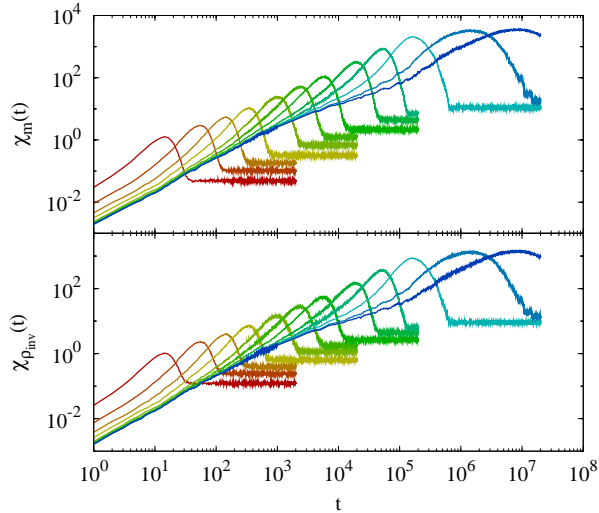}
  \caption{
  The dynamical susceptibility of the order parameter $\chi_{m}(t)$ and density of invisible states $\chi_{\rho_{\rm inv}}(t)$ of the ($3$,$27$)-state Potts model starting from the completely ordered state $\sigma_i(0)=1$ (for all $i$) for fixed different temperatures.
From left to right, the temperatures are $T=T_{\rm c}+0.5$, $T_{\rm c}+0.2$, $T_{\rm c}+0.1$, $T_{\rm c}+0.05$, $T_{\rm c}+0.02$, $T_{\rm c}+0.01$, $T_{\rm c}+0.005$, $T_{\rm c}+0.002$, $T_{\rm c}+0.001$, $T_{\rm c}+0.0005$, $T_{\rm c}+0.0002$, and $T_{\rm c}+0.0001$, where $T_{\rm c}=0.58513$\cite{Tanaka-2010d}.
  }
  \label{graph:dynchi}
 \end{center}
\end{figure}
Figure \ref{graph:dynchi} shows the dynamical susceptibility $\chi_m(t)$ and $\chi_{\rho_{\rm inv}}(t)$ starting from the completely ordered $\sigma_i(0)=1$ (for all $i$) state for fixed different temperatures of the ($3$,$27$)-state Potts model.
In Fig.~\ref{graph:dynchi}, the dynamical susceptibilities have a peak which indicates characteristic melting time.
Characteristic melting time defined from $\chi_m(t)$ and $\chi_{\rho_{\rm inv}}(t)$ are almost same.
Then we define $\tau_{\rm max}$ as the peak position of dynamical susceptibility of the order parameter $\chi_m(t)$.
As the temperature decreases, the peak height increases.
This result means that characteristic length scale becomes large as the temperature approaches to the transition point, since the peak height relates to characteristic length scale.
The characteristic time $\tau_{\rm max}$ also increases as the temperature decreases.
These dynamical susceptibilities are quantitatively similar with $\chi_4$ which is often used in analysis of glassy systems\cite{Krzakala-2010b}.

We also study temperature-dependency of $\tau_{\rm max}$ for the ($3$,$25$)-state Potts model and ($3$,$27$)-state Potts model to consider the effect of the number of invisible states.
\begin{figure}[t]
 \begin{center}
  \includegraphics[scale=0.65]{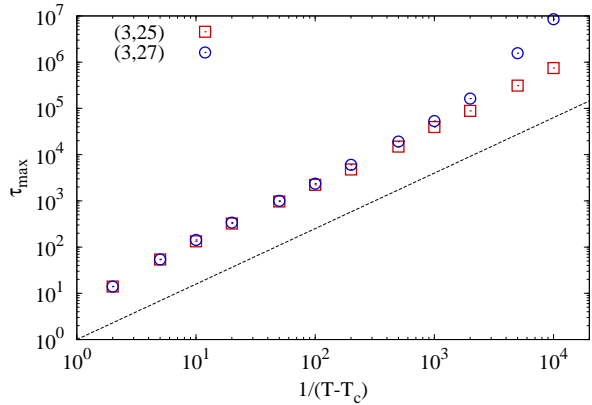}
  \caption{
  The peak position of the dynamical susceptibility of the order parameter $\tau_{\rm max}$ as a function of $1/(T-T_{\rm c})$, where $T_{\rm c}$ is the transition temperature.
  The red squares and blue circles indicate the case for the ($3$,$25$)-state Potts model and the ($3$,$27$)-state Potts model, respectively.
  Transition temperatures for the ($3$,$25$)-state Potts model and the ($3$,$27$)-state Potts model are $0.59630$ and $0.58513$, respectively\cite{Tanaka-2010d}.
  The dotted line indicates $1/(T-T_{\rm c})^{1.2}$.
  }
  \label{graph:tau}
 \end{center}
\end{figure}
Figure \ref{graph:tau} shows the peak position as a function of $1/(T-T_{\rm c})$, where $T_{\rm c}$ denotes transition temperature.
The transition temperatures for ($q$,$r$)=($3$,$25$) and ($3$,$27$) are $0.59630$ and $0.58513$, respectively\cite{Tanaka-2010d}.
In high temperature region, the peak position obeys $\tau_{\rm max} \propto 1/(T-T_{\rm c})^{1.2}$ whereas near the transition temperature, the characteristic time $\tau_{\rm max}$ becomes longer rather than $1/(T-T_{\rm c})^{1.2}$.
In high temperature region, behavior of $\tau_{\rm max}$ is the same both for ($3$,$25$)-state Potts model and for ($3$,$27$)-state Potts model.
However, there is difference near the transition temperature.
The characteristic time $\tau_{\rm max}$ for the ($3$,$27$)-state Potts model is larger than that for the ($3$,$25$)-state Potts model, since the energy barrier between the ordered state and paramagnetic state for the ($3$,$27$)-state Potts model is larger than that for the ($3$,$25$)-state Potts model\cite{Tamura-2010,Tanaka-2010d}.
From this result, we expect that $\tau_{\rm max}$ enlarges as the number of invisible state $r$ increases, since the latent heat increases as $r$ increases\cite{Tanaka-2010d}.
Thus, the Potts model with invisible states can be regarded as a standard model to analyze dynamic behavior of systems which exhibits a first-order phase transition.

\section{Conclusion and future perspective}

We studied dynamic properties of the Potts model with invisible states at fixed temperatures which are above the transition temperature.
In this paper, we focused on melting process which is an important nature of first-order phase transitions.
First, we considered the dynamical properties of the order parameter, density of invisible states, and the internal energy.
We found that there is two-step relaxation which is typical behavior in systems which exhibit a first-order phase transition and also glassy systems.
Next we calculated the dynamical susceptibility of the order parameter and density of invisible states.
As the temperature approaches the transition temperature, the peak height of the dynamical susceptibility grows.
Furthermore, we studied the relation between the peak position of the dynamical susceptibility of the order parameter $\tau_{\rm max}$ and temperature.
We found that the peak position obeys $\tau_{\rm max} \propto 1/(T-T_{\rm c})^{1.2}$ in high temperature region, whereas this relation breaks near the transition temperature.
The growth speed of $\tau_{\rm max}$ depends on the energy barrier between the ordered state and paramagnetic state.

In the Potts model with invisible states, we can control the energy barrier between the ordered state and paramagnetic state without changing the symmetry which breaks at the transition point.
We should study the relation between the energy barrier and the time-scale and length-scale in this model more carefully.
Mori and Nakada have studied dynamical properties in this model systematically to clarify whether there is new {\it universal} behavior near the first-order phase transition point\cite{Mori-inp}.

It is also important topic to clarify whether the Potts model with invisible states relates to strange type of first-order phase transition in two-dimensional frustrated systems. It will be reported elsewhere.

\ack
The authors are grateful to Naoki Kawashima, Jie Lou, Yoshiki Matsuda, Seiji Miyashita, Takashi Mori, Yohsuke Murase, Taro Nakada, Masayuki Ohzeki, and Eric Vincent for their valuable comments.
S.T. is partly supported by Grant-in-Aid for Young Scientists Start-up (21840021) from the JSPS, Grant-in-Aid for Scientific Research (B) (22340111), and the ``Open Research Center'' Project for Private Universities: matching fund subsidy from MEXT.
R.T. is partly supported by Global COE Program ``the Physical Sciences Frontier'', MEXT, Japan. 
The computation in the present work was performed on computers at the Supercomputer Center, 
Institute for Solid State Physics, University of Tokyo.

\end{document}